# Light Induced Electron-Phonon Scattering Mediated Resistive Switching in Nanostructured Nb Thin Film Superconductor

Shafaq Kazim[1], Alka Sharma[1,2], Sachin Yadav[1], Bikash Gajar[1,2], Lalit M. Joshi[1,2], Monu Mishra[2,3], Govind Gupta[2,3], Sudhir Husale[1,2], Anurag Gupta[1,2], Sangeeta Sahoo[1,2*] & V.N. Ojha[1,2]

[1]Time & Frequency and Electrical & Electronics Metrology, National Physical Laboratory, Council of Scientific and Industrial Research, Dr. K. S Krishnan Road, New Delhi-110012, India.

[2]Academy of Scientific and Innovative Research (AcSIR), National Physical Laboratory, Council of Scientific and Industrial Research, Dr. K. S Krishnan Road, New Delhi-110012, India.

[3]Advanced Materials & Devices Division, National Physical Laboratory, Council of Scientific and Industrial Research, Dr. K. S Krishnan Road, New Delhi-110012, India.

*Correspondence and requests for materials should be addressed to S.S. (sahoos@nplindia.org)

The elemental Nb is mainly investigated for its eminent superconducting properties. In contrary, we report of a relatively unexplored property, namely, its superior optoelectronic property in reduced dimension. We demonstrate here that nanostructured Nb thin films (NNFs), under optical illumination, behave as room temperature photo-switches and exhibit bolometric features below its superconducting critical temperature. Both photo-switch and superconducting bolometric behavior are monitored by its resistance change with light in visible and near infrared (NIR) wavelength range. Unlike the conventional photodetectors, the NNF devices switch to higher resistive states with light and the corresponding resistivity change is studied with thickness and grain size variations. At low temperature in its superconducting state, the light exposure shifts the superconducting transition towards lower temperature. The room temperature photon sensing nature of the NNF is explained by the photon assisted electron-phonon scattering mechanism while the low temperature light response is mainly related to the heat generation which essentially changes the effective temperature for the device and the device is capable of sensing a temperature difference of few tens of milli-kelvins. The observed photo-response on the transport properties of NNFs can be very important for future superconducting photon detectors, bolometers and phase slip based device applications.

The refractory metal Nb is famous for its superconducting properties and it is known to provide the highest critical temperature ($T_c$) (bulk $T_c = 9.2\ K$) for any elemental superconductor. Being one of the most used primary superconductors, Nb is well explored material in the field of superconducting photon detectors[1,2] and bolometric applications[3]. Besides, it exhibits various physical properties like high melting point, high thermal conductivity, high critical current density etc[4]. Since decades, dimensional effects of niobium on its physical properties like critical temperature, superconducting energy gap, critical field, coherence length, penetration depth etc. have been explored. These properties vary with the dimension of niobium particles/grains and/or films[5]. It is well known that nanomaterials can exist as various nanostructures such as quantum dots, nanowires, nanoparticles, etc. which play key role in upgrading their optoelectronic properties, mainly controlled by the quantum effect, as compared to their bulk counterpart[6][7]. For example, granular Nb thin films have been shown to act like Josephson junctions[5,8].

During 1970's the effects of light on conventional superconductors have been explored intensely[9]. The effect of light and transient photo-response on oxide based high-$T_c$ superconductors have also garnered a reasonable attention in the field of superconductivity based optoelectronic applications[10]. Here, we explore optoelectronic properties of NNFs at room temperature (RT) and at low temperature (LT), above and below the $T_c$. To the best of our knowledge, for the first time we show here that these NNFs strongly respond to the light exposure at RT and behave as photo-switches. In this study, current through the NNF based devices switch to high resistive state with light and returns back to a lower resistive state with removal of the light. Hence, the NNF based devices show familiar phenomenon of photoconductivity but in an anomalous fashion compared to that with the conventional photodetectors. As Nb is superconducting below its $T_c$, the NNFs can offer a platform to study the interaction between photoconductivity and superconductivity as that has been experimented in some other superconducting materials[11].

Negative photoconductivity (NPC) has been observed in semiconductors doped with III-V group elements[12], diamond thin films[13], metal nanoparticles[14], graphene[15], polycrystalline topological insulator[16] etc. In order to understand the mechanisms behind the observed NPC, different explanations like surface plasmon resonance (SPR)[6], electron-surface plasmon polaritons (SPPs) scattering and the related Joule heating effect at the grain boundaries[17], hot electrons trapping[18], energy gap opening[16] and trap level due to defects etc. are addressed in the literature. However, here we analyze various types of scattering mechanisms for conduction electrons in the presence of light to address the possible reason behind the room temperature NPC observed in the NNFs. We find photon assisted electron-phonon scattering is the main contribution to the NPC observed at room temperature. Further, under the illumination by visible (405 nm) and NIR (800 nm) light sources, the NNF device at its superconducting state switches sharply to a higher resistive state in a similar way how it responds to the light at RT. Interestingly, above the $T_c$ at 10 K in its metallic state, the NNF does not respond to any of the visible and NIR lights. And the low temperature NPC observed below the $T_c$ is mainly due to the light induced heating as it happens for the superconducting transition edge sensor (TES) based bolometric detectors[1]. These fascinating light sensing properties make these NNFs as promising candidates for the use in photo-switches, photodetectors and non-volatile memory with low power consumption based devices[18].

## Results:

We have fabricated NNFs based multi-terminal devices for the optoelectronic transport study at room temperature as well as at low temperature. A false coloured field emission scanning electron microscopy (FESEM) image along with the measurement scheme for a representative device is shown in Figure 1(a). Most of the measurements performed at RT were done in 2-probe geometry through the connection (3) and for LT, we used connections (1) & (2) in 4-probe geometry. However, to check the consistency with low temperature measurements with that at room temperature, we performed one set of measurements in 4-probe geometry at RT also and the results are presented in Figure S3 in the supporting material (SM). The first category of samples are made of ~100 nm thick NNF layer covered with ~10 nm thick Si capping layer using ultra high vacuum magnetron sputtering and the IVC, measured in two-wire geometry at room temperature under the dark condition, is shown in Figure 1(b).The linear IVC indicates the Ohmic nature of the contacts. The effect of the light illumination on the dc electrical transport is monitored by time dependent repetitive light 'Off' and 'On' cycles with halogen light, visible 532nm and 1064nm NIR lasers. With light exposure, current decreases for all the three light sources i.e., the device switches from a lower resistive state (LRS) to a higher resistive state (HRS) on illumination. The device current quickly attains its initial high dark current state after the removal of light. The resistive switching under the light exposure makes the device compatible as photodetector and photoswitch applications in the visible-to-NIR wavelength range. The change-in-current ($\Delta I$) is extracted by subtracting the dark current ($I_{Dark}$) from the current under light illumination ($I_{Light}$) as, $\Delta I = I_{Light} - I_{Dark}$[19]. Figure 1(c) exhibits the time-resolved $\Delta I$ measured at fixed bias voltage, $V_{DS}= 400\ mV$, with respect to the light 'On' and 'Off' cycles for the afore-mentioned three light sources. The current cycles show an overall downward slope with time for all the three lights with 532 nm laser having the maximum slope. The slope might be due to the Joule heating effect[17]. We have studied the bias dependent change-in-current ($\Delta I$) for $V_{DS}$ from 100 mV to 500 mV for all the three lights as shown in Figure 1(d). $\Delta I$ varies linearly with $V_{DS}$ indicating NNF as a good photodetector candidate[20].

In order to study the feasibility of using the NNF devices for promising photodetector based applications, we have measured $\Delta I$ while varying the optical power density (P) at constant $V_{DS} = 500\ mV$ and the same is shown in Figure 2(a). For P < 1.52 mW/cm$^2$, no significant $\Delta I$ is observed. The amplitude of $\Delta I$ increases linearly with P as shown in Figure 2(b). The linear power dependent $\Delta I$ indicates very low contribution of thermoelectric current[21] and the presence of low density trapped states[22]. Additionally in the left panel of Figure 2(b), we have presented the power dependent photo responsivity (R), measured at 500mV bias voltage, using equation[23,24], $R = \frac{I_{Photo}}{P \times A}$, where $A$ is the effective device area. A representative NNF device shows the maximum responsivity of ~2.65 A/W at 1.52 mW/cm$^2$ for 1064 nm NIR light. The responsivity is continuously decreasing with increasing P. It suggests that nanostructure based devices achieve maximum light absorption at low optical power density [25,26].

For an in-depth understanding, we have carried out three different types of experiments, namely, NNF devices (i) with and without Si capping layer, and variations in (ii) grain size and (iii) film thickness. Firstly, we investigate the role of Si capping layer on the observed NPC, we have prepared a batch of samples using the same growth conditions but without the capping layer. The film without capping showed a similar NPC type behavior with noticeably enhanced $\Delta I$ (~3 times greater) [Figure S1(c) in the SM] compared to that with Si capping layer presented in Figure 1. The suppression in $\Delta I$ for the device with capping layer is most likely due to the influence of the capping between the film surface and the light exposure and eventually the capping layer is influencing the light penetration to the NNF. Further, we have fabricated control samples with only sputter-deposited $a$-Si layer of thickness about 20 nm and have measured the IVCs under the exposure of light and eventually, we have observed positive photoconductivity in contrast to the observed NPC in the NNF samples. The results are shown in Figure S1 (e)-(f) in the SM. In order to know the role of the capping layer in formation of any possible silicide phase, we have performed x-ray photoelectron spectroscopy (XPS) of a reference sample having NNF covered with a Si capping layer and the results are shown in Figure S2 (c)-(d) in the SM. The detailed analysis of Nb 3d and Si 2p spectra indicate the major contribution of elemental phases of Nb and Si in the NNFs. However, the results also indicate a possible formation of Nb-Si based silicide phases for the NNFs with Si capping layer. Hence, Si capping can slightly influence the properties of Nb by forming its silicide phases in addition to its role as the protecting layer for the film from any adsorbed contamination due to ambient conditions. Finally, the dominant elemental phases of Nb and Si present in the NNFs clearly indicate that the observed NPC is mainly originated from NNF but there can be a minor influence from the possible Nb-Si based compound formed with a Si capping layer.

Further, we have fabricated NNFs with variations in grain sizes and thicknesses by high temperature growth and post-growth annealing in order to study the effect of grain size and hence grain boundaries on $\Delta I$ as reported also in the literature[27]. The change in grain sizes in the range of 25-135 nm is shown in Figure 3(a-d). The samples grown and annealed at 820ºC for 2 hours are having grain sizes ~3 times larger than the same for the samples grown at RT. Further increment in the grain sizes [Figure 3(c-d)] are achieved by increasing the deposition time while keeping the growth and annealing conditions unaltered. Correspondingly, $\Delta I$ increases for larger value of grain size as shown in Figure 3(e). The change in $\Delta I$ is noticeable, however, a direct correlation of the enhanced $\Delta I$ with the grain size is unclear from the measured data. One reason might be the thickness i.e., the thicknesses for first two devices are in the same range of ~ 100 nm (the red and black circles) whereas the other two devices having thicknesses of ~ 260 nm (the green triangles) and 480 nm (the blue triangles), respectively. Due to the change in thickness overall resistance changes and hence the effect of grain size on $\Delta I$ cannot be correlated in a simple straightforward way. We have observed reduced $\Delta I$ for thinner samples compared to that for the thicker films (Table S1 in the SM).

To assess the effects of the crystallinity and purity of the films on the observed NPC, we have characterized the films by using x-ray diffraction (XRD) pattern which is shown in figure 3(f). From the XRD data, most of the peaks confirmed the bcc structure of Nb. However, there are few peaks indicating the presence of niobium oxide phases observed for high temperature grown NNFs. The oxidation might occur at high temperature with the oxygen present in the $SiO_2$ /Si substrate. These oxide phases can play a crucial role in the observed NPC by acting as defects or impurities on the enhanced NPC for the high temperature grown NNFs.

The intriguing room temperature photo-response for the NNFs attracts the possibility to explore the material for the same at its superconducting state below $T_c$. At LT, superconducting nanowires are known to serve as potential candidates for single photon detectors[2] and in this regard how the presented NNFs respond to the light in their superconducting state is of interest for the following section. We have measured the temperature dependent resistance (R-T) characteristics and the related light effects for the same sample, presented in Figure 1(a), in 4 probe geometry. The main panel in Figure 4(a) represents the normal metal to superconductor (NM-SC) transition region from the R-T measurement along with its derivative, *dR/dT*, while cooling the device from RT to the liquid helium temperature. The R-T measurement for full temperature range is presented in the upper left inset of Figure 4(a). The $T_c$ is defined as the temperature where *dR/dT* becomes the maximum and it is ~ 8.58 K for the dark state as indicated by the red vertical line in Figure 4(a). The residual resistance ratio (RRR), defined as the ratio of the resistance measured at 300 K [$R_{T=300K}$] to the resistance measured at 10 K [$R_{T=10K}$], is 2.78 indicating the film is in dirty limit[28,29]. The transition width, *ΔT*, defined as the temperature extent between the resistance values related to 90% and 10% of the normal state resistance, is 0.25 K indicating a moderate switching for NM-SC transition. To investigate the light effect on the superconducting transition, we have measured the R-T characteristic under the light illumination with 800 nm NIR (the red curve) and 405 nm visible (the black curve) lights and the related transition curves are shown in Figure 4(b) along with the dark state R-T curve (the blue curve) for the comparison. We observe that the NM-SC transition shifts towards the lower temperature with increasing wave length. The $T_c$ values for the device under 405 nm and 800 nm lights are 8.49 K and 8.39 K, respectively. Under the light illumination, the shift of the transition towards the lower temperature is most likely originated from the light–induced heating. The effective temperature increases on light exposure and the transition shift to lower temperature. In order to have an estimate of the light induced change in temperature, *δT*, we have extracted the resistance values for a fixed temperature of 8.4 K (the green vertical line) from the measured three R-T curves with different light conditions shown in Figure 4(b). Resistance increases from its dark state value of ~ 0.23 Ω to ~ 0.48 Ω under 405nm light and to ~ 1.98 Ω with 800 nm light. The resistance values of 0.48 Ω and 1.98 Ω correspond to the temperature of 8.43 K and 8.57 K for the dark state R-T curve, respectively. Hence the *δT* for 405 nm light and 800 nm light are 0.03 K and 0.17 K, respectively. In other words, resistance increases due to energy absorption from light and the absorbed energy works as a form of heat increasing the effective temperature of the device[30]. Here, the device at its superconducting state is functioning like a TES based bolometer which can be used as a thermal detector[31]. For example, the device can sense about 30 mK change in temperature for 405 nm optical illumination by changing its resistance.

Further, we have measured time resolved resistance measurements with the laser lights 'On' and 'Off' conditions, shown in Figure 4(c), at a fixed temperature, $T = 8.4\ K$, which is just below the dark state $T_c$. The black and the red curves represent 405 nm and 800 nm lights, respectively. We observe reproducible sharp switching of resistance under the light exposure for both visible and NIR wavelengths and the resistance switches to a higher value when the light is on. The amplitude of the resistance change decreases with the wave length of the light exposure which is consistent with the shifting in transition presented in Figure 4(b). As we observe that the 'On'-state resistance value for the NIR light is the same (1.98 Ω) as that appeared in Figure 4(b) for the R-T curves under 800 nm light. However, for the 'Off'-state resistance values for both the 405 nm and 800 nm lights in the time dependent resistance measurements appeared as 0.15 Ω while in the R-T measurements in Figure 4(b), the same measures a resistance value of 0.23 Ω. The 'On' state resistance value (0.27 Ω) for 405 nm is also less than the value (0.48 Ω) obtained from the R-T measurements. The discrepancy in the resistance values, measured in two different measurement schemes as described in Figure 4(b) & (c), can be understood from the thermal fluctuations at the measured temperature (8.4 K) which is very close to $T_c$. Otherwise, a clear change in resistance due to light illumination is evident in both the types of measurements. Hence the reproducibility in the resistance change due to the light illumination and their corresponding temperature sensing by the NNF device demonstrate that the presented NNFs can be suitable for future superconducting bolometric applications[32]. However, at 10 K in the normal metallic state, the device does not respond to light illumination and the related time dependent resistance measured under the same light source is displayed in Figure 4(d). A more detailed study is needed to understand the mechanism behind the light response at low temperature for both superconducting and metallic states of the device.

## Discussions:

NPC have been observed in semiconducting nanostructures by localized trapped states[18], band gap opening in topological insulators[16], e-SPPs scattering in metallic nanowires[17,33], adsorbent molecule in graphene based composite systems[34], doping-induced formation of trions in $MoS_2$[35] etc. Here we address the origin of the observed NPC by calculating the dc resistivity from Drude's free electron model which deals with various scattering mechanisms experienced by conduction electrons while travelling under an electric field in the presence of light. First from the linear current-voltage characteristics, we calculate the change-in-current ($\Delta I$) with a fixed $V_{DS}$ as,

$$\Delta I = I_{Light} - I_{Dark} \propto \left(\frac{1}{\rho_{Light}} - \frac{1}{\rho_{Dark}}\right) \qquad (1)$$

Now from dark to light 'On' condition, the device switches from a lower to a higher resistive state, hence, $\rho_{Light} = \rho_{Dark} + \delta\rho$, where $\delta\rho$ is the light-induced resistivity. Therefore,

$$\frac{1}{\rho_{Light}} - \frac{1}{\rho_{Dark}} = \frac{1}{\rho_{Dark}+\delta\rho} - \frac{1}{\rho_{Dark}} = \frac{-\delta\rho}{\rho_{Dark}(\rho_{Dark}+\delta\rho)} = \frac{-\delta\rho}{\rho_{Dark}^2\left(1+\frac{\delta\rho}{\rho_{Dark}}\right)} \qquad (2)$$

For $\frac{\delta\rho}{\rho_{Dark}} \ll 1$ from equations (1) & (2),

$$I_{ph} \propto \frac{\delta\rho}{\rho_{Dark}^2} \qquad (3)$$

Here, $I$ and $\rho$ represent current and resistivity, respectively. The subscripts *Light* (*Dark*) represent the representative quantities under light 'On' ('Off') states, respectively. From eqn. (3) it is clear that the dark state resistivity ($\rho_{Dark}$) and the photo-induced resistivity ($\delta\rho$) control the $\Delta I$. Hence, the detailed contributions from the surface roughness along with the grain sizes and grain boundaries, film thickness and the light effect on the resistivity can be important for the understanding of the origin of the NPC observed in these NNF devices. For a particular device, $\rho_{Dark}$ is constant and therefore, $\Delta I$ depends only on $\delta\rho$.

Now the dc resistivity, $\rho_{dc}$, from Drude's free electron model can be written as,

$$\rho_{dc} = \rho = \frac{m_e^*}{n_e e^2 \tau} \propto \frac{1}{\tau} \qquad (4)$$

Where, $m_e^*$, $n_e$ and $e$ are the effective mass, density, and charge of free electron, respectively. The electron scattering time, $\tau$ depends mainly on temperature and the impurities along with their charge states present in the metal. Now, with constant $m_e^*$ and $n_e$ in the dark state, $\rho$ depends mainly on $\tau$ as evident in eq. (4). In a metal, electron-phonon, electron-electron and electron-impurity scattering mechanisms are the dominant bulk scattering contributions to the total scattering rate. Besides, surface contributions, controlled mainly by the surface roughness, grain sizes and grain boundaries, contribute significantly to the electron scattering rates[15,36]. Further, the scattering rates, $\left(\frac{1}{\tau^{bulk}}\right)$ and $\left(\frac{1}{\tau^{surf}}\right)$ can be added using Matthiessen's rule:

$$\frac{1}{\tau} = \frac{1}{\tau^{bulk}} + \frac{1}{\tau^{surf}} \qquad (5)$$

$$\frac{1}{\tau^{bulk}} = \frac{1}{\tau^{e-e}} + \frac{1}{\tau^{e-ph}} + \frac{1}{\tau^{e-imp}} \qquad (6)$$

$$\frac{1}{\tau} = \frac{1}{\tau^{e-e}} + \frac{1}{\tau^{e-ph}} + \frac{1}{\tau^{e-imp}} + \frac{1}{\tau^{surf}} \qquad (7)$$

At high temperature (higher than the Debye temperature) electron-phonon scattering is the dominant mechanism and all the other scattering mechanisms can be ignored. Whereas at very low temperature, the scattering is limited with the impurity and defect controlled electron-electron scattering and electron-phonon scattering would have negligible influence on the resistivity of the metal [37,38]. In the dark at very LT limit, the electron-electron scattering rate dominates and follows as, $\frac{1}{\tau^{e-e}} \propto T^2$ [39-41] and at RT, the dominant electron-phonon scattering rate scales linearly with temperature, i.e., $\frac{1}{\tau^{e-ph}} \propto T$ [40]. The electron-impurity scattering comes into play at very low temperature and $\frac{1}{\tau^{e-imp}}$ can be considered as temperature independent[37,42]. The surface scattering time, $\frac{1}{\tau^{surf}} = g_s \frac{v_F}{L_{eff}}$, where $g_s, v_F, L_{eff}$ correspond to the surface properties, the Fermi velocity and the effective dimension of nanostructured particles, respectively[43]. The $L_{eff}$ for spherical particles represents the diameter (D) while for nanowires, $L_{eff} \approx (LD)^{1/2}$ with $L$ and $D$ as the length and the diameter of the nanowire, respectively[44].

In the presence of light with frequency, ω, the electron-electron scattering time can be related with the effective temperature ($T'$) and ω as, $\frac{1}{\tau^{e-e}} \propto (T'^2 + \omega^2)$ [40,45 46,47]. The electron-phonon scattering rate, $\frac{1}{\tau^{e-ph}} \propto T'$ [46,48] and the electron-impurity scattering rate, $\frac{1}{\tau^{e-imp}}$ remain almost unaltered with the optical frequency [45]. Here, $T' = T + \delta T$ with $\delta T$ as the change in temperature due to light-induced heating[33,49,50]. However, the contribution from the surface scattering in the presence of light will have an additional term related to the scattering of conduction electron with the photo-induced SPPs[17,33,49]. The electron-SPPs scattering rate can be expressed as, $\tau_{light}^{e-SPP} \propto P\lambda_{mfp}^2$ with $P$ as the laser power and $\lambda_{mfp}$ is the mean free path[17,51,52]. Now,

$$\frac{1}{\tau_{Light}^{surf}} = \frac{1}{\tau_{Dark}^{surf}} + \frac{1}{\tau_{Light}^{e-SPP}} = g_s \frac{v_F}{L_{eff}} + \gamma_s P\lambda_{mfp}^2 \cdot \frac{\lambda_m}{L_{eff}} \qquad (8)$$

Where, $\gamma_s$ contains the information about the surface properties and the distribution of SPPs along the interface between the metallic film and the ambient air[53]. $\lambda_m$ is the penetration depth. Combining eq. (8) with eq. (7), one can obtain the resistivity under light illumination by knowing the total scattering rate which eventually depends on the effective temperature ($T'$), excitation frequency (ω) and the power ($P$) of the light source. Here for simplicity, the electronic mass and the electron density are assumed to have no effect of the light exposure.

We now consider the room temperature photo-response for the device presented in Figure 1. A very little variation in the amplitude of $\Delta I$ for 1064 nm (29 mW/cm$^2$) and 532 nm (32 mW/cm$^2$) lasers is observed which can be attributed to the slightly higher laser power density for 532 nm. No such strong effect of the wavelength is observed and hence, we can neglect the frequency dependence of $\frac{1}{\tau^{e-e}}$ in the case of the studied two wavelengths, *viz.*, 532 nm and 1064 nm. The surface morphology shown in Figure 3(a) for this device represents a granular structure having grain sizes in the range of 20-30 nm which can excite SPs while coupled to light. Hence, the room temperature NPC might be originated from the interaction between electrons and photo-excited SPPs. However, the same device does not respond to light in its metallic state at *T = 10 K* where in principle, e-SPPs scattering should be much stronger mainly due to the negligible influence of electron-phonon scattering and secondly, the electronic mean free path ($\lambda_{mfp}$) increases at low temperature and the e-SPPs scattering rate depends on the square of $\lambda_{mfp}$ [eq. (8)]. Hence, the observed NPC for the metallic NNF is not related to e-SPPs scattering. Further at LT, the total scattering rate is dominated by the impurity controlled electron-electron scattering rate for which a quadratic dependence on temperature and frequency has been reported[40,45-47]. However, we do not observe any reasonable change in the device resistance under the illumination of light in the NIR and visible range for 800 nm and 405 nm, respectively, at 10 K [Figure 4 (d)]. Light induced heating causing a change in the temperature $\delta T$ can be important to consider as the origin of the NPC at RT. $\delta T$ relates to the heat dissipation mechanism which depends on the thermal conductivity of films and the substrate and the interface heat conduction between the film and the substrate (related to the Kapitza resistance)[54]. To investigate the detail heat dissipation mechanism through the substrate one needs to determine the individual contribution from the substrate, films and the interface. In the present study, NNFs are in strongly dirty limit and the RRR value is very low (RRR ~ 2.8) indicating not so significant change in thermal conductivity at 10 K compared to the same at room temperature[55]. Besides, the thermal conductivity is reduced for $SiO_2$/Si substrate at 10 K from its room temperature value[56]. Therefore, to the simplest assumption, we can ignore the light induced heating effect as the origin of the NPC observed at RT compared to the presence of strong electron-phonon scattering at RT. Therefore, photon assisted electron-phonon scattering mechanism might be the origin for the observed NPC for the metallic state of the NNF as reported in other materials also[45,57].

In summary, at room temperature NNFs with different grain sizes and thicknesses show negative photoconductivity which gets enhanced for bigger grain sizes and thicker films. Further, the low temperature transport measurements under light illumination offer strong photo-response in superconducting state but above the $T_c$, no significant light effect is observed for the NNF devices.

## Methods:

The NNFs were deposited on highly doped p-type Si (100) substrate topped with thermally grown $SiO_2$ layer of thickness 300 nm as the dielectric spacer. Prior to loading into the vacuum chamber, the substrates were cleaned in acetone and isopropanol for 15 minutes each in a sonication bath followed by an oxygen plasma cleaning for 10 minutes. Before deposition the substrates were pre-heated under high vacuum at ~ 800 °C for 30 minutes in order to remove any residual organic and tenacious contaminations. The chamber was evacuated to less than $3.4 \times 10^{-9}$ Torr and the substrates were brought back to the room temperature before starting the sputtering. Here we report for two types of growth conditions depending on the substrate temperature during the deposition. For the first type, the substrate temperature was kept at room temperature while for the second type the same was fixed at ~ 820°C during the sputtering of the Nb film. The Nb target used in this study was with 99.99% purity. The sputtering was done in an Ar (99.9999% purity) atmosphere of ~ $3.7 \times 10^{-3}$ mbar. For this study, the film thickness was varied from 100 nm – 500 nm and for some of the films we deposited a thin Si capping layer of ~ 10 nm thickness. We fabricated the contact leads of Au (100 nm)/Ti (5 nm) using sputtering too. Shadow masks were used to define the Nb film based channel and the electrical contacts. For low temperature 4-probe measurements [Figure 1(a)], connections (1) & (2) are used as the current sourcing and voltage measuring, respectively. Whereas, room temperature two probe measurements are performed through the connection (3). The distance between the voltage leads [connection (2)] is ~ 1.3mm. However, for room temperature measurements the distance between two leads and the width of the channel are 106μm and 86μm, respectively.

The room temperature transport measurements were carried out using a Keithley 2634b source measure unit in a probe-station from Cascade Microtech with shield enclosure. We have used (i) halogen light, (ii) 532 nm laser light in the region (power ~ 32 mW/cm$^2$), and (iii) 1064 nm laser light in NIR range with (power ~ 29 mW/cm$^2$) as the light sources for illuminating the devices. The low temperature measurements were done in a SQUID magnetometer (Quantum Design) with a homebuilt fiber coupled optical excitation using wavelength specific filters in the visible and NIR region[58]. The low temperature measurements were done under the lights with 405 nm (~1.5-2 mW/cm$^2$) wavelength in the visible and 800 nm (~ 6 mW/cm$^2$) in the NIR regions.

## ACKNOWLEDGEMENTS:

This work is supported by the network project- AQuaRIUS (project No.: PSC-0110). We are indebted to Dr. K.K. Maurya for the XRD characterization and Mr. Sandeep Singh for his help with AFM imaging. We are thankful to Mr. M. B. Chhetri for his technical help in the laboratory. S.K. is thankful to Prof. M. A. Wahab and Miss Hana Khan for critical reading of the manuscript. S.K. acknowledges the financial support for her research internship from CSIR-NPL. B.G. acknowledges the JRF fellowship from UGC-RGNF.## Author Contributions:

S.K, S.Y., B.G. and S.S fabricated the devices & analyzed the results. A.S., S.K. and S.H performed room temperature photoconductivity measurements and participated in the data analysis. L.M.J. and A.G performed the low temperature measurements coupled with light. M.M. & G.G carried out the XPS measurements. S.K. and S.S. wrote the manuscript with the help from S.H., S.Y., B.G. and A.S. The project is supervised by S.S., S.H. and V.N.O. All the authors read and reviewed the manuscript thoroughly.

# Additional information

Competing financial interests: The authors declare no competing financial interests.

# Figure Captions:

**Figure 1: Room temperature photo-response of a NNF device.** (a) The device geometry and the measurement connections. Room temperature two-probe measurements are done through the terminals marked as (2) and low temperature 4-probe measurements are with (1) & (3) connections. (b) Current –voltage characteristic of NNF at room temperature in dark condition. Inset: A magnified SEM image of the NNF device measured in two probe geometry. (c) Time dependent measurement of $\Delta I$ for repetitive cycles of light 'On' and 'Off' states under halogen light, 532 nm and 1064 nm lasers for $V_{DS} = 400\ mV$. (d) Bias dependent $\Delta I$ for the studied three

light sources. The scattering points represent the experimental points whereas, the solid lines are the linear fits.

**Figure 2: Power dependence for 1064 nm NIR light. (a)** A set of time dependent change-in-current ($\Delta I$) curves measured with different optical power density (1.52 mW/cm2 – 29 mW/cm2) at $V_{DS}$ = *500 mV* **(b)** Variation of change-in-current ($\Delta I$) and responsivity with optical power densities. The blue circles represent the extracted $\Delta I$ with respect to different power densities and the same is fitted linearly as shown by the red line.

**Figure 3: Morphological and structural analysis of NNFs having variations in the grain sizes and thicknesses and their dependence on the measured change-in-current ($\Delta I$).** (a-d) AFM topography images representing the variations in the grain sizes for differently grown NNFs. The size varies in the range between 20 nm -140 nm. The colour bars represent the height variations of 0-6.6 nm, 0-9.4 nm, 0-12 nm, and 0-18 nm for (a), (b), (c), and (d) respectively. (e) Bias dependent change-in-current ($\Delta I$) for the presented 4 samples along with their linear fits shown by the solid lines. (f) HRXRD spectra of 3 out of the 4 reference samples. The oxide phases of niobium appear for the high temperature grown samples.

**Figure 4: Effects of light on the low temperature transport properties.** (a) R-T measurements in the dark condition. The $T_c$, defined as the temperature corresponding to the maximum of $dR/dT$, is shown by the green vertical line indicating $T_c = 8.58\ K$. Inset: The full scale R-T data measured from room temperature 6K. (b) R-T measurements with 405 nm and 800 nm wavelength lights along with the dark state R-T curve. (c) Real time response of the device resistance to light 'On' and 'Off' states at 8.4 K in the superconducting state of the device. The red and black curves indicate the resistance switching for 800 nm 405 nm lights, respectively. (d) Effects of the light on devices measured at 10 K in its metallic state. The color shades indicate the light 'On' and 'Off' states. No significant change in resistance observed at 10 K

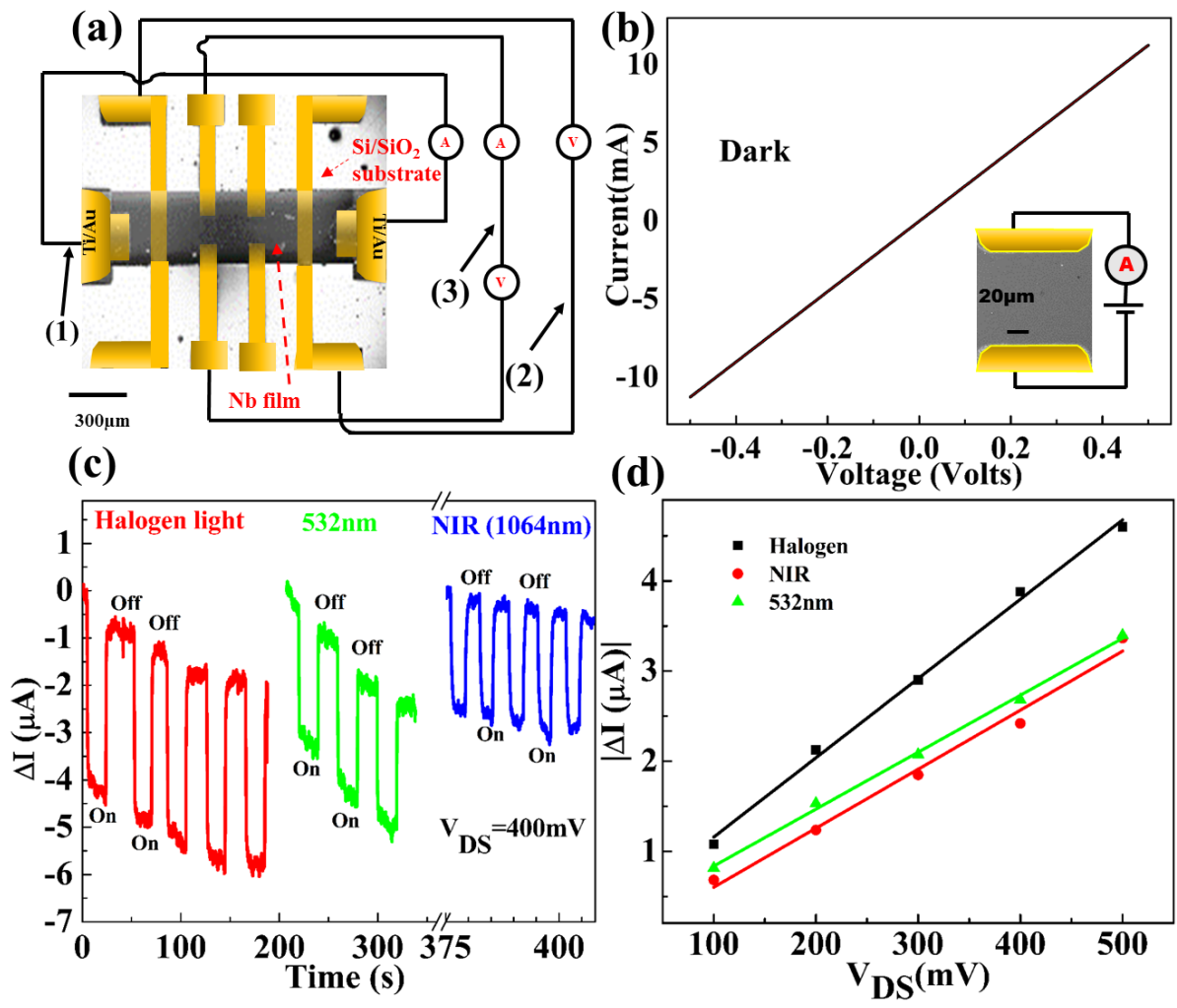

**Figure-1**

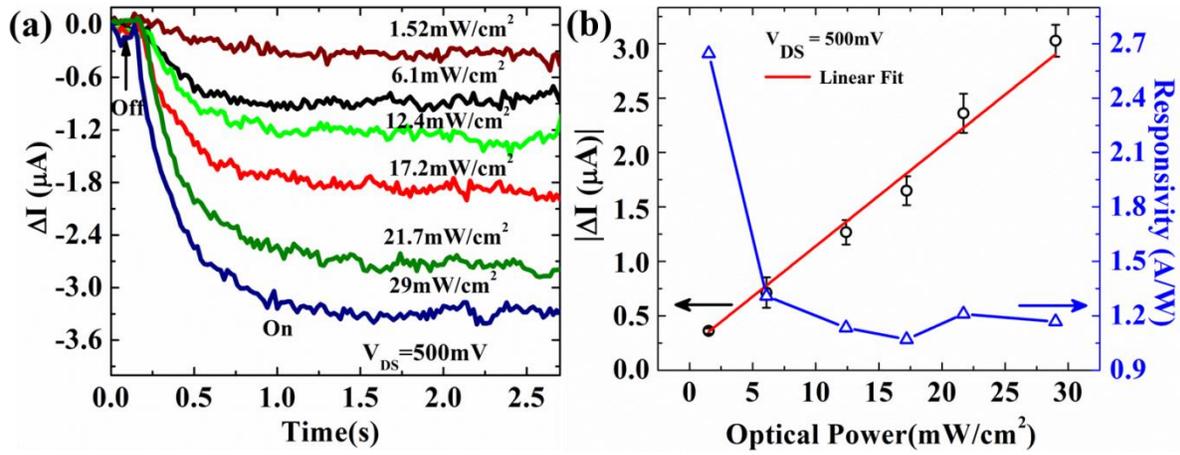

**Figure-2**

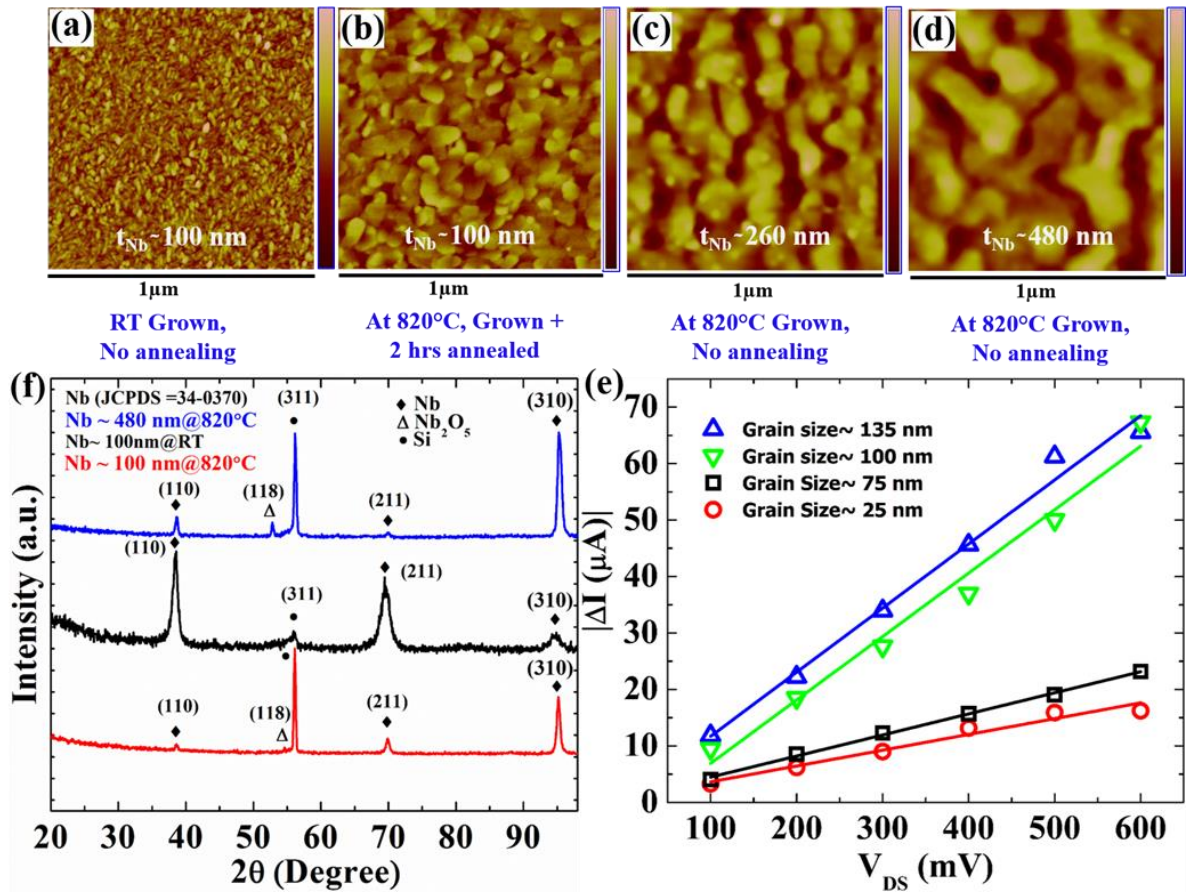

**Figure-3**

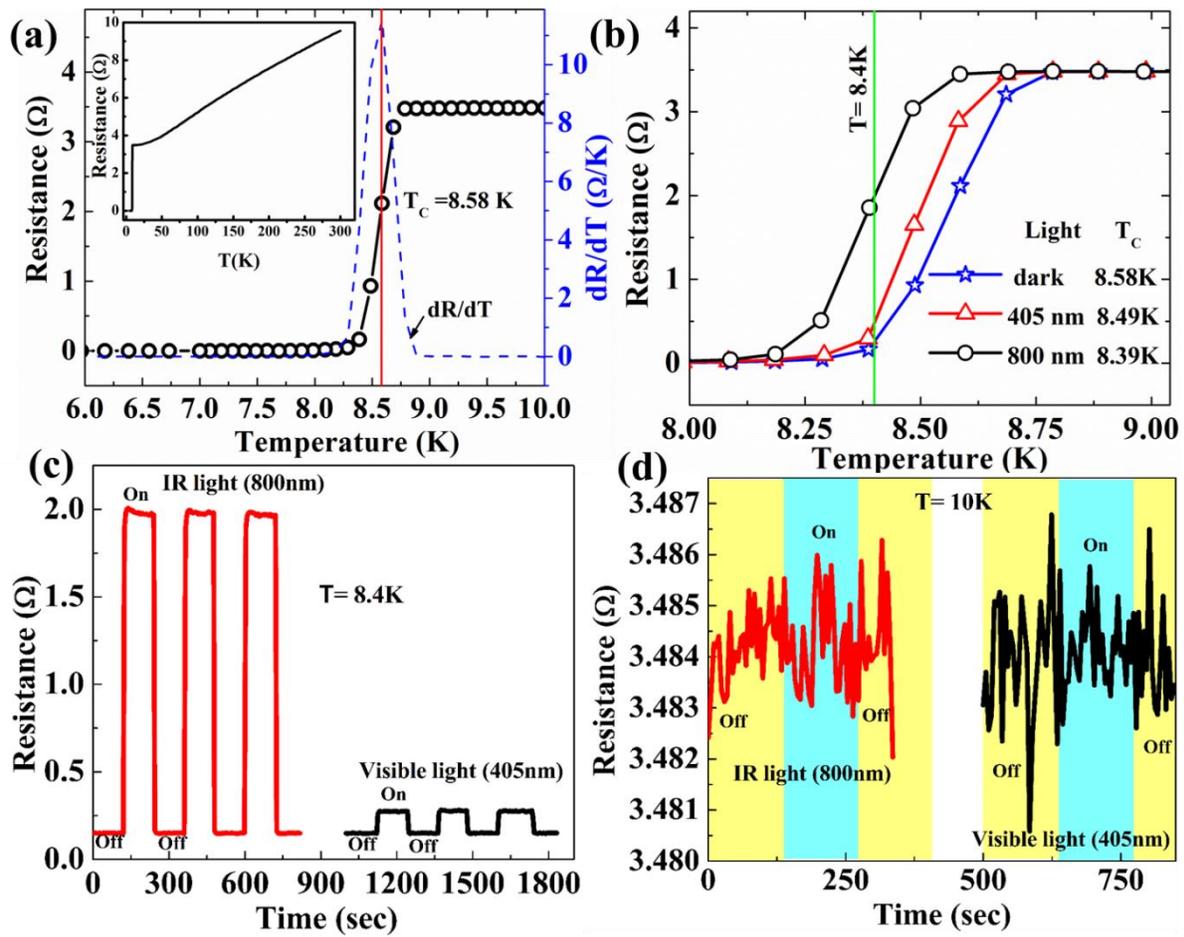

Figure-4

Supporting Material

# Light Induced Electron-Phonon Scattering Mediated Resistive Switching in Nanostructured Nb Thin Film Superconductor

*Shafaq Kazim[1], Alka Sharma[1,2], Sachin Yadav[1], Bikash Gajar[1,2], Lalit M. Joshi[1,2], Monu Mishra[2,3], Govind Gupta[2,3], Sudhir Husale[1,2], Anurag Gupta[1,2], Sangeeta Sahoo[1,2\*] and V.N. Ojha[1,2]*

[1]*Time & Frequency and Electrical & Electronics Metrology, National Physical Laboratory, Council of Scientific and Industrial Research, Dr. K. S Krishnan Road, New Delhi-110012, India.*

[2]*Academy of Scientific and Innovative Research (AcSIR), National Physical Laboratory, Council of Scientific and Industrial Research, Dr. K. S Krishnan Road, New Delhi-110012, India.*

[3]*Advanced Materials & Devices Division, National Physical Laboratory, Council of Scientific and Industrial Research, Dr. K. S Krishnan Road, New Delhi-110012, India.*

*E-mail: sahoos@nplindia.org

# Contents in Supporting Material:-

1. Control experiment to evaluate the role of Si capping layer on the observed NPC

2. Cross-sectional FESEM and XPS study: observation of interface and study the elemental phases of Nb and Si in the Si/Nb bilayer film.

3. Representative 4-probe measurements at room temperature for a similar sample presented in Figure 1 in the main manuscript.

4. Summary of various sample parameters used in this work in a tabular form

### 1. Control experiment to evaluate the role of Si capping layer on the observed NPC:

In order to study the role of Si capping layer used as the protective layer on the devices, we have carried out the optoelectronic measurements on two different NNFs based devices, viz., with and without the capping layer. The AFM topography images representing surface morphology are shown in Figure S1 (a) & (b). The related photo-response with time dependent $\Delta I$ measurements are displayed in Figure S1 (c) & (d). As evident from the AFM morphology, no significant changes appear in the overall morphology and the grain sizes for both the types. However, we observe an enhanced change-in-current ($\Delta I$) for the device having no capping layer. This indicates initially that the capping layer acts as a barrier for the film to get direct exposure to the light and the effects result in reduced $\Delta I$. Further to evaluate the role of silicon capping in originating NPC, we have performed similar optoelectronic measurements at room temperature using 532 nm laser and halogen light sources on only *a*-Si films with no contribution from NNFs. In result, we find linear IV characteristics in dark as well as under light conditions and along with the positive photoresponse from Si capping layer as shown in Figure S1 (e) & (f).

This confirmed that NPC is originating from the NNF, whereas Si capping layer is acting as protecting layer from the adsorbed contamination from the environment.

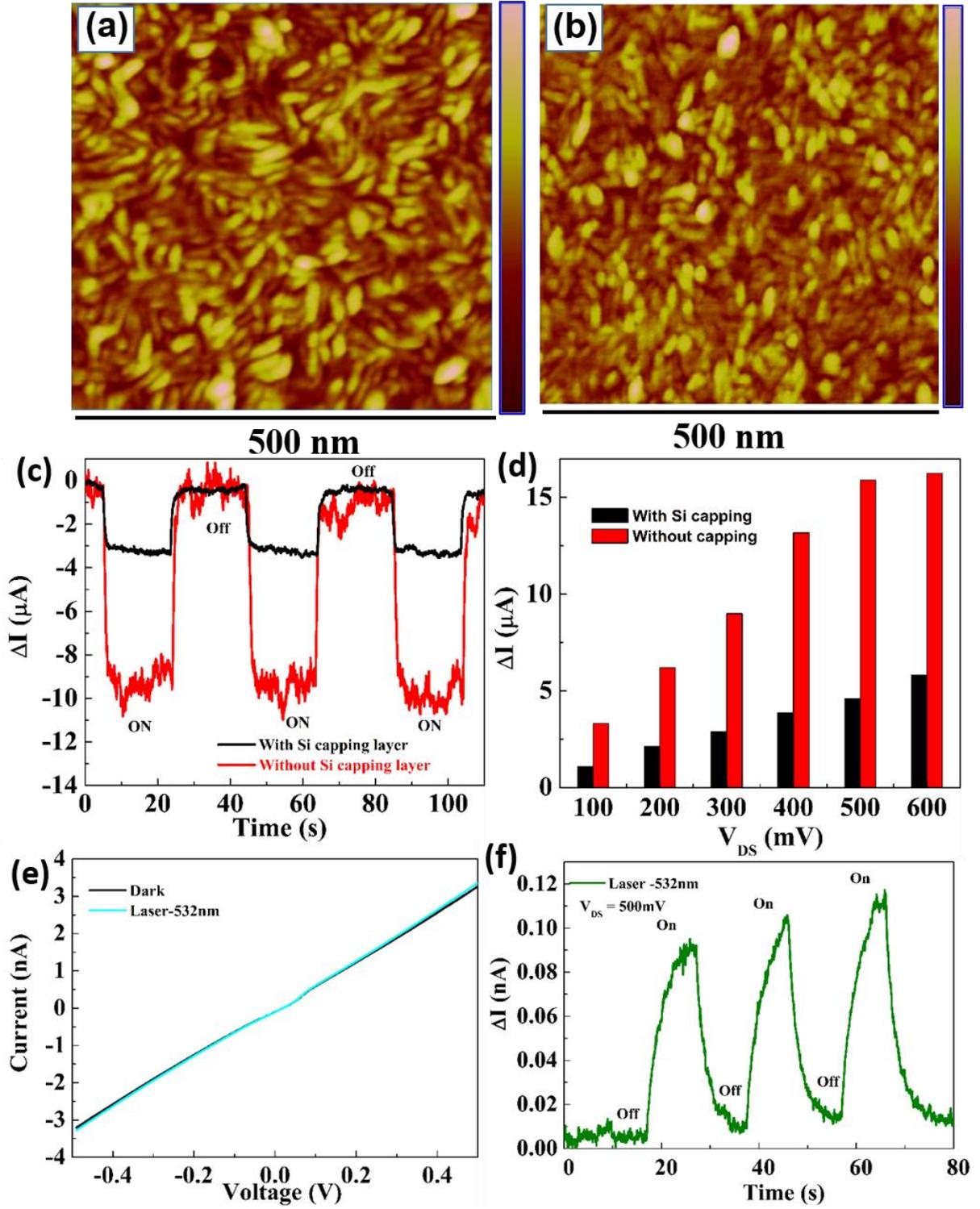

*Figure S1:* *AFM morphology of NNFs (a) without & (b) with Si capping layer. The grains look almost same with grain sizes in the range of 20-30 nm. The colour bars represent the height*

variations from 0 to 5 nm in (a) and 0 to 6.6 nm in (b). (c) The corresponding real-time measurement of $\Delta I$ for light 'Off' and 'On' cycles. (d) The comparison chart for the bias dependent $\Delta I$ for the afore-mentioned two types of devices. (e) IV characteristic for a- Si film having same device geometry under both the conditions. (f) Change in current with time at 500 mV bias voltage for a-Si thin film and shows positive photo-response.

**2. Cross-sectional FESEM and XPS study: observation of interface and study the elemental phases of Nb and Si in the Si/Nb bilayer film:**

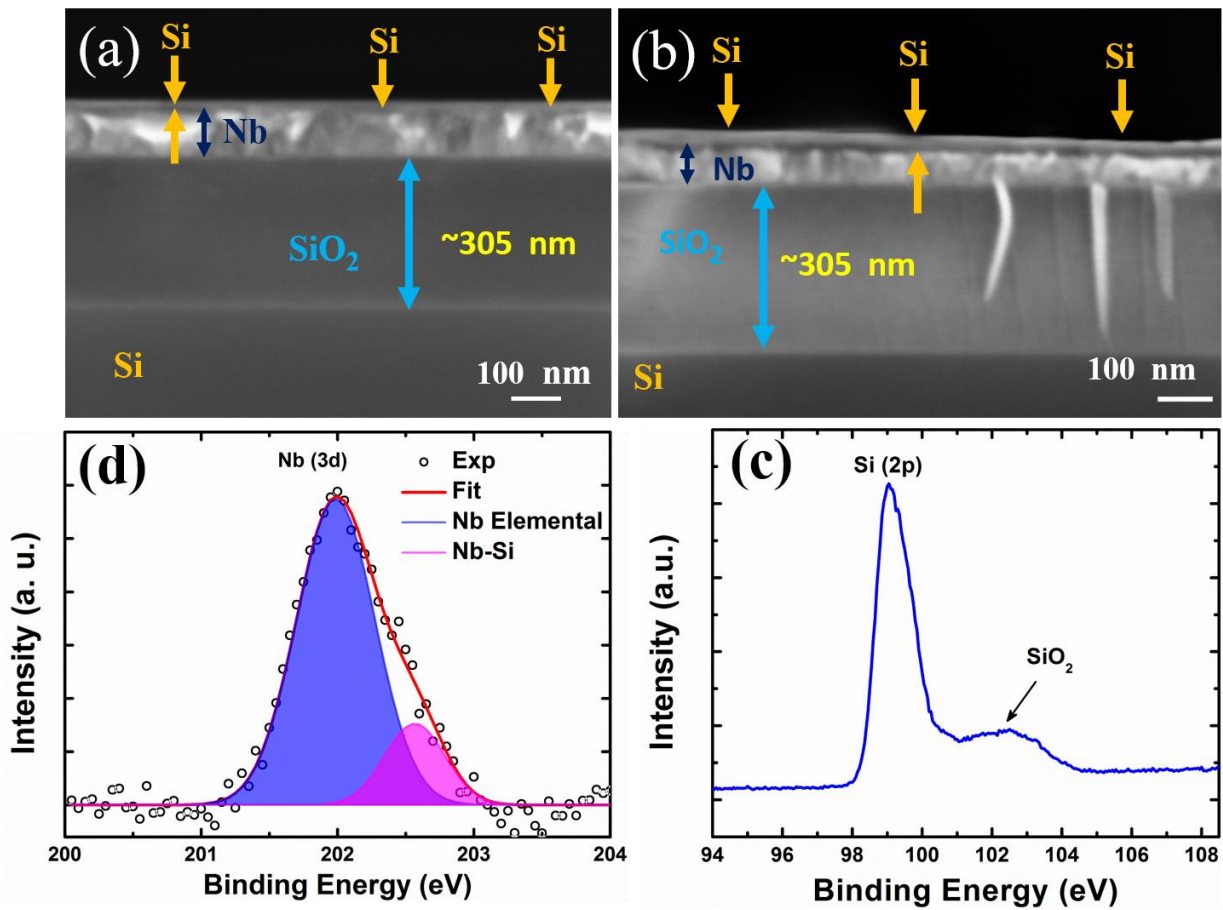

Figure S2: **(a) & (b) FESEM Cross-Sectional view for Nb thin film with Si capping layer on top of it. XPS data presenting (c) Nb 3d and (d) Si 2p spectra for the film with Si capping layer.**

FESEM Cross-sectional view gives a clear indication about the interface between Nb and Si thin film deposited on $SiO_2$/Si substrate as marked by different arrows in Figure S2 (a) & (b). For further confirmation, we have performed XPS studies on similar type of thin film samples, and the results are shown in Figure S2 (c) & (d). As there was a capping of Si thin layer on top of the NNFs, soft $Ar^+$ ions sputtering (500 eV, 5 minutes) was performed to remove few nanometres of Si capping layer. The binding energy position of Nb $3d_{5/2}$ appears at 202.0 eV which confirms the dominant presence of elemental Nb. The peak is further de-convoluted to analyse the exact chemical state. It is observed that the Nb $3d_{5/2}$ peak consists of two components corresponding to elemental Nb (85%) and Nb-Si (15%) with a chemical shift of 0.6 eV[1]. The slight presence of the Nb-Si phase can also have influence on the amplitude of the observed NPC. However, from the XPS results it is evident that the major contribution is from elemental Nb in the studied NNFs. Further, the Si (2p) core level is located at 99.1 eV and reveals the elemental phase of silicon. The $SiO_2$ formation is witnessed by the shoulder peak observed at ~103 eV.

[1] Matthew, J.A.D., Morton, S.A., Walker, C.G.H. and Beamson, G. Auger parameter studies of amorphous NbSi. J. Phys. D: Appl. Phys. **28,** 1702 (1995).

## 3. Representative 4-probe measurements at room temperature for a similar sample presented in Figure 1 in the main manuscript:

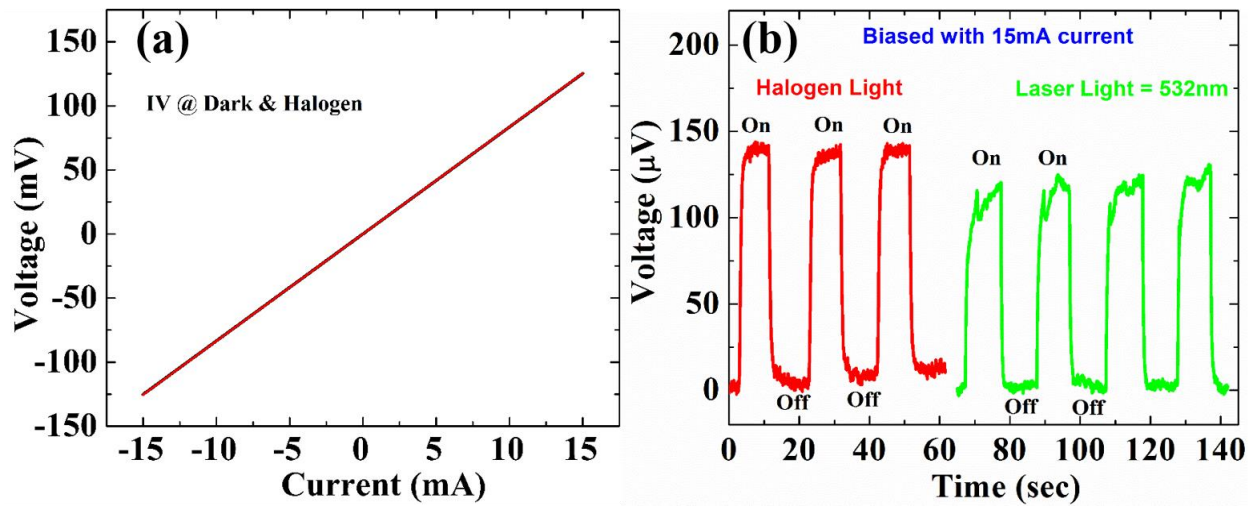

Figure S3: **The room temperature optoelectronic measurements in four probe Geometry: (a) IVC of a NNF device under the dark and illuminated conditions. IVC appears to be linear and follows Ohm's law. (b) The time dependent voltage characteristics with a constant bias current of 15 mA. The red and green curves correspond the switching behaviour of film with halogen and 532nm laser light respectively.**

In order to compare the device geometry used for performing low temperature measurements on same samples, we have measured the photoresponse by using four-probe measurements at room temperature under halogen and laser light (532 nm) as shown in Figure S3, where current-voltage characteristics (IVC) show linear behavior under dark as well as light conditions and time dependent voltage characteristic biased at 15 mA shows stable switching under light illumination in the four probe measurements and that is similar to two-probe measurements at room temperature.

**4. Summary of various sample parameters used in this work in a tabular form:**

| Figures related to AFM | Growth Condition | Thickness (nm) | Grain size (nm) | ΔI (µA) @ 300mV | ΔI (µA) @ 400mV |
|---|---|---|---|---|---|
| Figure-3(a) | RT grown, No annealing | 100nm | 25nm | 8.99825 | 13.17065 |
| Figure-3(b) | 820°C Grown and 2hrs annealed | 100nm | 75nm | 12.2816 | 15.675 |
| Figure-3(c) | 820°C Grown, No annealing | 260nm | 100nm | 27.61 | 50.02053 |
| Figure-3(d) | 820°C Grown, No annealing | 480nm | 135nm | 33.97771 | 61.26364 |